\documentclass[a4paper,fleqn]{cas-dc}

\usepackage[numbers]{natbib}

\def\tsc#1{\csdef{#1}{\textsc{\lowercase{#1}}\xspace}}
\tsc{WGM}
\tsc{QE}
\tsc{EP}
\tsc{PMS}
\tsc{BEC}
\tsc{DE}

\begin{document}
\let\WriteBookmarks\relax
\def\floatpagepagefraction{1}
\def\textpagefraction{.001}

\shorttitle{Decision Support System to triage of liver trauma}

\shortauthors{Ali Jamali et~al.}

\title[mode = title]{Decision Support System to triage of liver trauma}                      



%
\author[1]{Ali Jamali}






\affiliation[1]{organization={CSE and IT Department, Shiraz University},
    city={Shiraz},
    country={Iran}}

\author[2]{Azadeh Nazemi}

\affiliation[2]{organization={Perth, Western Australia},
    country={Australia}}

\author[3]{Ashkan Sami}[orcid=0000-0002-0023-9543]
\ead{a.sami@napier.ac.uk}
\cormark[1]

\affiliation[3]{organization={Computer Science, Edinburgh Napier University},
    city={Edinburgh},
    country={United Kingdom}}

\author[4]{Rosemina Bahrololoom}

\affiliation[4]{organization={Microbiology clinic research centre of medical science, Shiraz University of Medical Sciences},
    city={Shiraz},
    country={Iran}}

\author[5]{Shahram Paydar}

\affiliation[5]{organization={Department of Surgery, School of Medicine, Shiraz University of Medical Sciences},
    city={Shiraz},
    country={Iran}}

\author[6]{Alireza Shakibafar}

\affiliation[6]{organization={Taba Medical Imaging centre and Shiraz University of Medical Sciences},
    city={Shiraz},
    country={Iran}}

\cortext[cor1]{Corresponding author}



\begin{abstract}
Trauma significantly impacts global health, accounting for over 5 million deaths annually, which is comparable to mortality rates from diseases such as tuberculosis, AIDS, and malaria. In Iran, the financial repercussions of road traffic accidents represent approximately 2\% of the nation's Gross National Product each year. Bleeding is the leading cause of mortality in trauma patients within the first 24 hours following an injury, making rapid diagnosis and assessment of severity crucial. Trauma patients require comprehensive scans of all organs, generating a large volume of data. Evaluating CT images for the entire body is time-consuming and requires significant expertise, underscoring the need for efficient time management in diagnosis. Efficient diagnostic processes can significantly reduce treatment costs and decrease the likelihood of secondary complications. In this context, the development of a reliable Decision Support System (DSS) for trauma triage, particularly focused on the abdominal area, is vital. This paper presents a novel method for detecting liver bleeding and lacerations using CT scans, utilising the GAN Pix2Pix translation model. The effectiveness of the method is quantified by Dice score metrics, with the model achieving an accuracy of 97\% for liver bleeding and 93\% for liver laceration detection. These results represent a notable improvement over current state-of-the-art technologies. By facilitating quicker diagnoses, this system could contribute to more timely interventions, potentially saving lives and further reducing healthcare costs associated with prolonged diagnostic procedures.

\end{abstract}


\maketitle

\section{Introduction}

Trauma is a major global health issue, leading to over 5 million deaths annually, similar to the toll from diseases like tuberculosis, AIDS, and malaria \citep{b14}. Annually, road accidents cause more than 1.2 million deaths globally \citep{b15}. Despite affecting all nations, over 90\% of trauma-related deaths occur in low- and middle-income countries \citep{b14} \citep{b16}. Trauma injuries are among the top 10 global causes of death \citep{b17}, with predictions that they will rise to the fourth leading cause of death by 2030 \citep{b18}. The 2019 Global Burden of Disease study highlights Iran's notably high road traffic accident mortality rate, with a significant increase from 2009 to 2019 \citep{b19}. In Iran, the cost of road traffic accidents amounts to about 2\% of the GNP annually \citep{b20}.

The early detection of abdominal trauma, specifically liver bleeding with or without other organs bleeding, is crucial for patient survival and effective treatment. Despite the critical importance of timely diagnosis, the availability of sophisticated diagnostic systems in emergency departments is often limited or non-existent. This gap in trauma care technology has motivated the current study, which aims to develop a software system for the accurate diagnosis of bleeding and liver lesions in trauma settings. The urgency of this need underscores the potential of artificial intelligence (AI) systems to significantly improve clinical outcomes by enabling faster and more precise trauma care triage.

AI-based CT scans can significantly enhance the efficacy and accuracy of triaging patients in a referral trauma centre. Time is critical in trauma care, often referred to as the “golden hour,” where rapid assessment and intervention can drastically improve patient outcomes. The process of triaging a trauma patient can be time-consuming, especially during high patient influx, such as earthquakes or large-scale traffic accidents. The following capabilities are aimed to be built:

\textbf{Timely and Precise Diagnoses:} AI can rapidly evaluate CT scans to assist healthcare providers in making timely and precise diagnoses. This is particularly important during high workload periods, where the likelihood of misdiagnosis increases due to the large number of patients. Automated segmentation of bleeding areas can rapidly identify the volume of bleeding and provide an accurate indication of the severity of the injuries, allowing for immediate and appropriate medical interventions.

\textbf{Reduction in Diagnostic Errors:} Trauma centres often handle large numbers of patients simultaneously, increasing the likelihood of minor injuries going unnoticed by radiologists. In cases of multi-organ trauma, radiologists may miss lacerations and haemorrhages due to artefacts in CT images or the complexity of the injuries. An accurate automated AI system can assist in identifying these subtle injuries, reducing the risk of diagnostic errors.

\textbf{Enhanced Detection of Multi-Organ Trauma:} In severe accidents, patients often suffer from multi-organ trauma. Automated segmentation of lacerated organs and bleeding areas can help identify and differentiate injuries across multiple organs, providing a comprehensive assessment essential for effective treatment planning. This capability is particularly beneficial in scenarios where radiologists may be overwhelmed or when their expertise is not immediately available. 
 The AI-based segmentation system should generalise across various internal organs and detect bleeding regions regardless of the specific organ involved. This generalisation is crucial as it ensures the system’s applicability to a wide range of trauma scenarios, enhancing its utility in diverse clinical settings.

This research introduces the initial milestones achieved in creating a comprehensive AI-driven system to address these concerns in trauma care triage processes. Initially, we faced the daunting challenge of not finding open-source online data regarding traumatic patients, especially for abdominal traumatic injuries, which our research aimed to study. To fine-tune advanced AI models, a similar dataset is necessary.  
We set up experiments to show that our algorithm can accurately detect the liver and its bleeding area.  The algorithm can easily extend to show lacerations in other internal organs. Since we did not have a large corpus of trauma data, we utilised 20 annotated abdominal DICOM CT scans of malignancies for finetuning. Our study employs the Pix2Pix Generative Adversarial Network (GAN) architecture for accurate organ segmentation and haemorrhage identification. This system not only aims to improve precision in multi-organ segmentation but also to streamline and enhance the efficiency of emergency medical services through automated diagnostic processes.

Central to our study is the adaptability of the Pix2Pix GAN model, initially developed for liver segmentation. This model has been extended to include the segmentation of critical organs such as the kidneys and spleen, applying transfer learning techniques to leverage prior knowledge from liver imaging. Employing a modified UNet architecture, our system is trained on an enriched dataset, including annotated images of additional organs, which enhances its capability to identify and differentiate various organ characteristics crucial for accurate segmentation.

The subsequent sections of this paper are organised as follows: Section 2 reviews the background literature to provide context and identify gaps in the current literature concerning medical image processing and AI applications in trauma care. Section 3 describes the proposed method, including dataset preparation, model selection, and the technical specifics of system development. Section 4 outlines the system development steps from conceptualisation to deployment. Section 5 discusses the proposed framework in detail and presents the clinical validation and potential of this AI system to improve diagnostic and treatment strategies in emergency medical care. Finally, Section 6 concludes the paper by providing a summary of findings and potential future research directions.

\section{Related work}

In this section, we discuss recent advances in medical imaging for liver and kidney segmentation, providing a chronological overview. We then present segmentation algorithms used for image segmentation in general, along with their strengths, shortcomings, and potential pitfalls.

\subsection{Recent Advances in Medical Imaging}
In 2007, Taourel et al. focused on vascular emergencies in liver trauma, examining vascular complications and elementary lesions shown by CT. They developed a grading system based on CT features and analysed its utility and limitations \citep{b1}. Yan et al. (2010) proposed an automatic method for kidney segmentation from 2D abdominal CT images using a combination of medical anatomical knowledge and traditional image processing techniques. This method improved upon earlier segmentation techniques by providing a more efficient and accurate solution \citep{b2}. 

Farzaneh et al. (2017) proposed a fully automated Bayesian-based method for 3D segmentation of the liver, achieving high similarity coefficients and demonstrating its effectiveness \citep{b3}. Vivanti et al. proposed a method for the automatic detection of new tumors and quantification of tumor burden in liver CT scans using a global convolutional neural network classifier, achieving a high true positive rate for new tumor detection \citep{b4}. Qiangguo Jin et al. proposed RA-UNet, an attention-aware 3D hybrid segmentation method for liver and tumor segmentation, combining low-level and high-level features to enhance accuracy \citep{b5}. Dreizin et al. conducted a retrospective study for automatic blunt hepatic injury detection using a multiscale deep learning algorithm, involving voxel-wise measurements and the use of Classification and Regression Trees (CART) \citep{b6}.

In 2021, Conze et al. utilized a GAN network for medical image segmentation of the liver, kidneys, and spleen, achieving a high accuracy rate of approximately 97\% \citep{b7}. Ayoob et al. emphasised the importance of imaging in diagnosing traumatic injuries to the pancreas and the necessity of an AI-based support system for early diagnosis using CT images \citep{b8}. Harry et al. proposed an AI-based approach for assessing body composition on routine abdominal CT scans and predicting mortality in pancreatic cancer, using a neural network to quantify tissue components and explore their relationship with mortality \citep{b9}. Dreizin et al. evaluated a multi-scale deep learning algorithm for the quantitative visualisation and measurement of traumatic hemoperitoneum, comparing its diagnostic performance with traditional methods \citep{b10}. Yang et al. introduced an improved unsupervised learning-based framework for multi-organ registration on 3D abdominal CT images, achieving accurate results suitable for real-time applications \citep{b10}.

In 2022, Farzaneh et al. proposed an end-to-end pipeline for calculating the percentage of liver parenchyma disrupted by trauma using deep convolutional neural networks. Their model achieved high Dice scores for both liver parenchyma and trauma regions, showcasing a significant advancement in trauma imaging technology \citep{b11}.

The methods described above, while pioneering and instrumental in advancing medical image analysis, exhibit certain limitations stemming from challenges such as high variability in patient anatomy, difficulty in distinguishing between similar tissues, reliance on high-quality imaging data, and the need for extensive labelled datasets for supervised learning approaches. Traditional image processing techniques and even advanced machine learning models may struggle with the nuances of medical imaging, such as capturing fine details in organ boundaries or accurately segmenting organs. Although Bayesian approaches and neural networks are powerful, they require significant computational resources and may not generalise well across diverse datasets without extensive tuning.

\subsection{Image Segmentation Algorithms}
Generally, image segmentation methods use traditional or deep learning approaches with various drawbacks:

\textbf{Thresholding:} It often produces poor results in images with varying lighting conditions or overlapping intensity values among different objects. It is not suitable for complex images where background and foreground cannot be easily separated based on intensity alone \citep{21}.

\textbf{Region-Based Segmentation:} Sensitive to noise, and the choice of seed points can significantly affect the outcome. Region splitting and merging are computationally expensive and can result in over-segmentation if not carefully managed \citep{22}.

\textbf{Edge Detection:} Susceptible to noise, and can miss important edges if not properly tuned. Edge detection methods also do not provide regions but only boundaries, requiring additional steps to segment the image \citep{23}.

\textbf{K-means Clustering:} Can converge to local minima and is sensitive to the initial choice of centroids.

\textbf{Fuzzy C-means Clustering:} Computationally intensive and sensitive to noise and outliers \citep{24}.

\textbf{Watershed Segmentation:} Leads to over-segmentation, especially in the presence of noise and minor gradient variations \citep{25}.

\textbf{Active Contour Model:} Very sensitive to the initial contour placement and may converge to local minima, requiring careful parameter tuning \citep{26}.

\textbf{Markov Random Fields (MRF):} The optimisation process can be computationally intensive, and performance heavily depends on the correct setting of parameters and the model's configuration \citep{27}.

\textbf{Deep Learning Methods:}

\textbf{Convolutions Neural Networks (CNNs):} Requires a large amount of labelled training data and computational resources. CNNs can be prone to overfitting, especially on small datasets without proper regularisation \citep{28}.

\textbf{U-Net:} May overfit to features specific to the training set unless sufficiently regularised \citep{29}.

\textbf{Fully Convolutional Networks (FCN):} Struggle with generalising to new, unseen images that differ significantly from the training data \citep{30}.

These disadvantages highlight the limitations and challenges of applying traditional segmentation methods to real-world medical images, emphasising the need for careful selection and tuning based on the specific requirements and constraints of each application. While publicly available datasets on cancer exist, we could not find open-source or publicly available datasets on abdominal injuries, specifically liver lacerations. Improving outcomes could benefit from fine-tuning specialised datasets for lacerations. It would be advantageous to use an algorithm refined on datasets involving malignant conditions but versatile enough for segmenting various organs in trauma patients. This research aims to address these issues.

\section{Data and Method}

\subsection{Data}
This research utilises two sets of datasets. 1) 3DIRCAD is an open-source CT images of cancer patients and 2) a set of trauma patients CT images from Shahid Rajaee Hospital. 
\subsubsection{Standard 3DIRCAD}
The 3DIRCAD (3D Image Reconstruction for Comparison of Algorithm Database) \citep{b13} is a specialized resource that offers a collection of medically annotated images, mainly in the form of DICOM files. These files are commonly used for picture archiving and communication in medical fields. The dataset primarily supports anatomical education and medical simulation.

The primary dataset, known as 3D-IRCAD, consists of 20 computed tomography (CT) scans, evenly split with 10 scans from female patients and 10 from male patients. The 3DIRCAD dataset includes 2823 images from 20 patients and has been extensively used for training models in liver segmentation.  The summary of the dataset is shown in Tables 1 and 2. Specifically, the dataset includes 1153 liver mask annotations, with 467 images also featuring annotations for tumors or cysts.


\begin{table}[width=.9\linewidth,cols=4,pos=h]
\caption{3DIRCAD Dataset Summary}\label{tbl1}
\begin{tabular*}{\tblwidth}{@{} LLLL@{} }
\toprule
Category & Count\\
\midrule
Total Patients & 20 \\
Total Images & 2823 \\
Liver Masks & 1153 \\
Tumors or Cysts & 467 \\
\bottomrule
\end{tabular*}
\end{table}


This research also utilized 3DIRCAD masks for multi-organ segmentation of the kidneys and spleen.



\begin{table}[width=.9\linewidth,cols=4,pos=h]
\caption{Kidney and Spleen Segmentation in 3DIRCAD}\label{tbl2}
\begin{tabular*}{\tblwidth}{@{} LLLL@{} }
\toprule
Category & Count\\
\midrule
Left Kidney & 432 \\
Right Kidney & 388 \\
Spleen & 397 \\
\bottomrule
\end{tabular*}
\end{table}

\subsubsection{Hospital Dataset in the Trauma Department (Rajaee Dataset)}
This dataset was collected from Shahid Rajaee (Emtiaz) Trauma Centre of the Medical Sciences University of Shiraz, Iran. The CT scan data was obtained from a Siemens-Somatom 16-slice device.

The Rajaee dataset includes imaging from 20 patients, with 1979 images used for liver segmentation. Of these, 632 images contain liver masks, and 310 include lacerations, focusing on liver condition analysis. These annotations were meticulously crafted using ImageJ, an open-source software for scientific image processing and analysis. The annotations precisely delineate liver areas and identify liver lacerations or bleeding, with corresponding masks created for each.

Two radiologists from the hospital conducted the initial image labeling, generating these masks over 20 days, dedicating about 4 hours per patient case, totaling approximately 80 hours of labeling work. Subsequently, a senior radiologist reviewed and revised these masks, dedicating about 10 minutes per case, adding up to an additional 3 hours for the entire verification process.


\begin{table}[width=.9\linewidth,cols=4,pos=h]
\caption{Rajaee Dataset Summary}\label{tbl3}
\begin{tabular*}{\tblwidth}{@{} LLLL@{} }
\toprule
Category & Count\\
\midrule
Total Patients & 20 \\
Total Images & 1979 \\
Liver Masks & 632 \\
Liver Laceration Masks & 310 \\
\bottomrule
\end{tabular*}
\end{table}


Researchers interested in advancing the field can request access to this dataset, which is provided by medical professionals and radiologists at the target hospital. This dataset is particularly valuable for developing and refining algorithms intended for the analysis of liver-related injuries and conditions.

\subsection{Method}

\subsubsection{Pix2Pix GAN}
This research employs the Pix2Pix Generative Adversarial Network (GAN), introduced by Phillip Isola et al. in 2017, to enhance medical image segmentation, specifically for liver laceration detection. Pix2Pix outperforms prior methods by achieving higher accuracy, as measured by the Dice score \citep{b12}.

The framework consists of two competing networks: a generator that produces images indistinguishable from real images, and a discriminator that distinguishes between real and generated images. This adversarial process ensures the generator produces highly realistic and detailed images.

Pix2Pix is well-suited for conditional image generation tasks, where outputs are directly conditioned on inputs. Conditional image generation refers to the process of creating images based on specific input images. Instead of generating images from random noise, the model uses an input image to guide the creation of the output image. This ensures that the output image is directly related to the input image. This capability is crucial for medical imaging tasks such as tumour detection or liver laceration segmentation, enabling the model to learn complex mappings from input CT images to output segmentations.
Pix2Pix's ability to generate detailed, high-quality images helps in the accurate identification of medical conditions, making it a robust tool for applications where precise and high-quality segmentation is crucial. The adversarial training compels the generator to focus on details and textures, essential for identifying subtle medical anomalies. It operates effectively by learning from paired data, where each input corresponds directly to an output segmentation map. It adapts to different tasks by modifying the training dataset and network architecture.

Figures 1 and 2 illustrate the architecture of the generator and discriminator, respectively.

\begin{figure*}[h]
  \includegraphics[width=\textwidth,height=6cm]{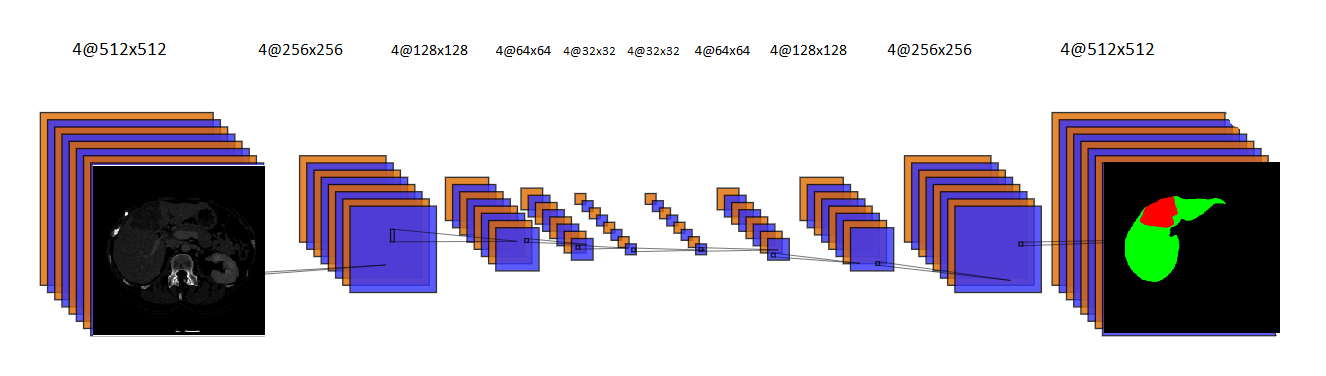}
  \caption{The architecture of the generator.}
\end{figure*}

\begin{figure*}[h]
  \includegraphics[width=\textwidth,height=6cm]{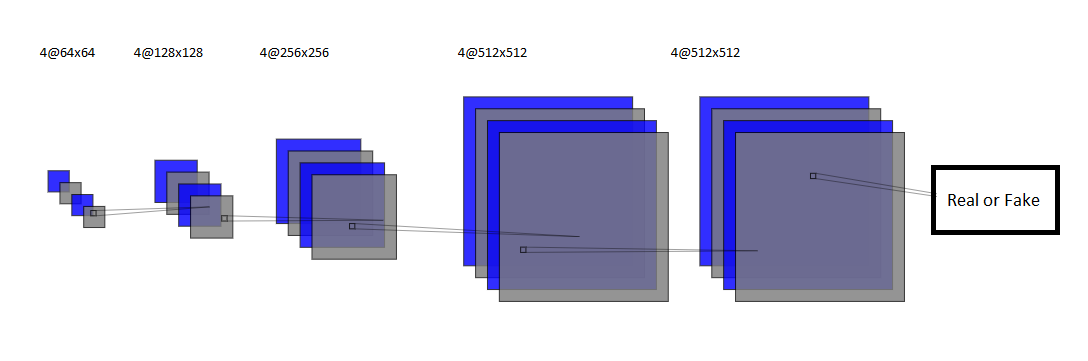}
  \caption{The architecture of the discriminator.}
\end{figure*}

The generator model updates in response to the discriminator's feedback. The generator tries to deceive the discriminator, while the discriminator attempts to recognize artificial images. Since output generation depends on input, the Pix2Pix model is called a conditional GAN (cGAN). The generator uses adversarial loss to generate realistic images in the target domain and updates through the loss of generated images compared to the expected output image (L1 loss).

\subsubsection{How to Develop and Train a Pix2Pix Model}
The Keras deep learning framework is used to build the discriminator and generator models, though the original implementation was based on Torch. The final generated image is a $256*256$ pixels color image. The discriminator, or conditional classifier, is a deep convolutional neural network that accepts the source image and the target image as input and predicts the probability of the image being fake or real.

The PatchGAN used in this model defines the relationship between the model's output and the number of pixels in the input images. It assigns each output prediction to a $70*70$ square or patch of the input image. The PatchGAN advantage applies to different sizes, whether larger or smaller than $256*256$ pixels.

The output size of the model has a linear relation to the input image. Input images contain concatenated images of sources and targets. Therefore, following the basic paper, the $70*70$ PatchGAN discriminator model is required. The model takes two input images joined together and predicts a patch output. Binary cross-entropy optimisation is used in this model. Based on the author's recommendation, weighting updating reduces the speed of discriminator changes.

The tangent hyperbolic (tanh) activation function in the output layer ensures pixel values in the generated image are in the range [-1,1]. Generator updates minimise the discriminator loss for predicted generated images labelled "real," leading to more realistic images. Additionally, the generator minimises the L1 loss between the generated and target images by measuring the mean absolute error. It updates the weighted sum of the adversarial and L1 losses. Finally, the generator accepts the source image, and the discriminator evaluates the generator's output against the related target.

The discriminator updates directly, and weights are used in the compound model with non-trainable labels. The updating process involves two targets (real and fake) with two loss functions: cross-entropy and L1. Images of source and target are concatenated, and then all images are compressed into a NumPy array for training. Generally, in the GAN model, an equilibrium is found between the generator and discriminator models instead of converging, making the decision to quit the training process complicated. The training process involves a constant number of iterations. Each epoch happens in one training iteration, and the generator model is saved after each ten epochs. Using GPU hardware is recommended for training purposes. Discriminator and generator losses are reported in each iteration.

\subsection{Preprocessing Steps Before Training}

\begin{itemize}
  \item \textbf{Organize the Data:}
  \begin{itemize}
    \item Collect all patient images into a folder named \texttt{images}.
    \item Collect all liver segmentation masks into a folder named \texttt{labels}.
  \end{itemize}

  \item \textbf{Convert DICOM to JPEG:}
  \begin{itemize}
    \item Convert the DICOM images to JPEG format to facilitate easier handling and processing during model training.
  \end{itemize}

  \item \textbf{Colorize CT Slides:}
  \begin{itemize}
    \item Transform grayscale CT images into full-color images to enhance the visibility of subtle features.
  \end{itemize}

  \item \textbf{Pixel-wise Label Mask Adjustment:}
  \begin{itemize}
    \item For each colorized image, pair it with its corresponding binary (black and white) liver mask to create a grayscale liver mask. This ensures each liver area retains its specific segmentation details in grayscale.
  \end{itemize}
\end{itemize}

Figure 3 illustrates the result of preprocessing.

\begin{figure}
\includegraphics[width=\columnwidth,height=4cm]{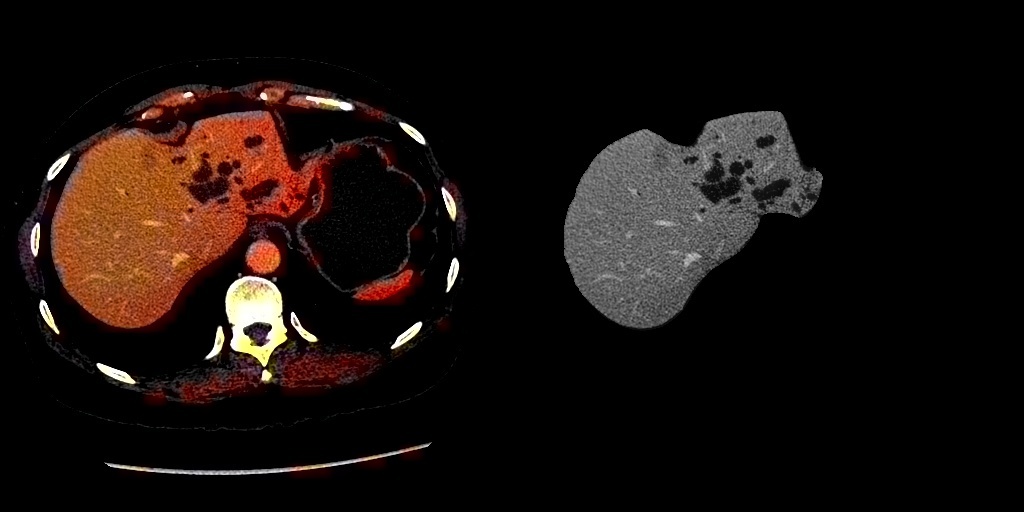}
   \caption{Preprocessing result}
\end{figure}
\section{Experimental Results}

The purpose of this research was to address the time-consuming challenges associated with abdominal injury detection at the Shahid Rajaee (Emtiaz) Trauma Centre of the Medical Sciences University of Shiraz, Iran. We show our algorithm can detect lacerations with high Dice scores.  Auto-
mated segmentation of lacerated organs and bleeding areas identified and differentiated injuries across multiple organs
provides a comprehensive assessment essential for effective
treatment planning.  The detection process relied on CT scan data obtained from a Siemens-Somatom 16-slice device. The proposed method involves the use of a modified UNet as a generator and a CNN as a discriminator.

In this section, we discuss three sets of results. Initially, we detail outcomes from training and testing on the 3DIRCAD standard dataset, which was divided using a 60/40 split for training and testing purposes, respectively. The models exhibited high accuracy with this dataset. Subsequently, we explore results from training and testing on our annotated dataset derived from real-world settings. Lastly, we demonstrate the applicability of our model in real-world settings where training data may not be present by training it on the standard dataset and testing it on Shahid Rajaee Hospital data.

\subsection{Experimental Results on Standard Dataset}
In evaluating the performance of segmentation algorithms, the Dice score was employed to measure the number of common pixels over the total number of pixels in an image. For the initial modeling and cross-validation, 3DIRCAD was utilized. The developed Pix2Pix image translation network achieved Dice scores of over 97\% and 93\% for liver and liver tumor detection on the testing subset, respectively.


\begin{table}[width=.9\linewidth,cols=4,pos=h]
\caption{3DIRCAD Data Model Performance}\label{tbl4}
\begin{tabular*}{\tblwidth}{@{} LLLL@{} }
\toprule
Segmentation & Train/Test Split & Dice Score (\%)\\
\midrule
Liver & 60/40 & 97 \\
Liver Tumor & 60/40 & 93 \\
\bottomrule
\end{tabular*}
\end{table}

Separate models for kidneys and spleen were trained and tested, achieving high Dice scores: 97.8\% (left kidney), 95.4\% (right kidney), and 91.3\% (spleen).


\begin{table}[width=.9\linewidth,cols=4,pos=h]
\caption{3DIRCAD Models for Kidneys and Spleen}\label{tbl5}
\begin{tabular*}{\tblwidth}{@{} LLLL@{} }
\toprule
Organ & Dice Score (\%)\\
\midrule
Left Kidney & 97.8 \\
Right Kidney & 95.4 \\
Spleen & 91.3 \\
\bottomrule
\end{tabular*}
\end{table}


\subsection{Experimental Results on Shahid Rajaee Dataset}
Medical professionals and radiologists supported this research by providing 20 cases of annotated CT scans featuring liver lacerations or bleeding conditions. The performance of the trained models using the real-world dataset is shown in Table 6.


\begin{table}[width=.9\linewidth,cols=4,pos=h]
\caption{Rajaee Data Models Performance}\label{tbl6}
\begin{tabular*}{\tblwidth}{@{} LLLL@{} }
\toprule
Organ & Dice Score (\%)\\
\midrule
Liver & 60/40 & 96.3 \\
Liver Laceration & 60/40 & 90 \\
\bottomrule
\end{tabular*}
\end{table}


\subsection{Experimental Results on Real-World Data While Trained on Another Dataset}
The performance of applying 3DIRCAD models for liver and liver tumor segmentation on the Rajaee dataset for liver and liver laceration is shown in Table 7.


\begin{table}[width=.9\linewidth,cols=4,pos=h]
\caption{Model Application to External Data}\label{tbl7}
\begin{tabular*}{\tblwidth}{@{} LLLL@{} }
\toprule
Segmentation & Dice Score (\%)\\
\midrule
Liver & 80 \\
Laceration & 74 \\
\bottomrule
\end{tabular*}
\end{table}


This research surpassed the state-of-the-art presented by Farzaneh et al. in 2022, particularly in segmenting the laceration portion of the image.

The initial model, trained using the standard dataset, was applied to the local dataset and demonstrated the high accuracy of the proposed method. When testing the model on real-world data, it achieved superior results in the segmentation of laceration compared to what Farzaneh et al. achieved on standard data. The summary of the Dice scores for each of these experiments is shown in Tables 8 and 9.

As illustrated in Table 8, the model, when tested on the Rajaee dataset without previous exposure, achieved a Dice score of 80\% for liver segmentation and 74\% for laceration segmentation. Considering the differing settings of the imaging instruments and the fact that the 3DIRCAD dataset is tailored for cancer studies, while the Rajaee dataset focuses on abdominal injuries, this performance highlights the versatility and applicability of our proposed method to broader conditions in the liver and other related issues. Confirming our hypothesis, Table 9 represents the performance of applying models trained on 3DIRCAD for kidney segmentation on the Rajaee dataset, achieving 75\% and 73\% Dice scores.

Figures 4 to 9 illustrate samples of outputs obtained by applying our model on Rajaee data without prior exposure. Figures 4 and 6 show liver segmentations. Figure 5 shows liver laceration segmentation of the Rajaee dataset by our model trained on malignant tumors. Figures 7, 8, and 9 show the left kidney, right kidney, and spleen segmentation of real-world data by our model while trained on standard data, respectively.


\begin{table}[width=.9\linewidth,cols=4,pos=h]
\caption{Dice Score for 3DIRCAD Dataset Model and Rajaee Real-World Data}\label{tbl8}
\begin{tabular*}{\tblwidth}{@{} LLLL@{} }
\toprule
Research & Liver & Lesion\\
\midrule
2017-Vivanti et al.-3DIRCAD & 86 & 72 \\
2022-Farzaneh et al.-3DIRCAD & 96.13 & 51.21 \\
This Research-3DIRCAD & 97 & 93 \\
This Research-Rajaee Real Data & 96.3 & 90 \\
This Research-3DIRCAD/Rajaee Test & 80 & 74 \\
\bottomrule
\end{tabular*}
\end{table}


\begin{table}[width=.9\linewidth,cols=4,pos=h]
\caption{Model Testing Results for Kidney Segmentation}\label{tbl9}
\begin{tabular*}{\tblwidth}{@{} LLLL@{} }
\toprule
Organ & Dice Score (\%)\\
\midrule
Left Kidney & 75 \\
Right Kidney & 73 \\
\bottomrule
\end{tabular*}
\end{table}

\textbf{Note:} A real test dataset was not available at the time of this report, precluding the performance evaluation of the trained spleen segmentation model in a clinical setting.

\begin{figure}
\includegraphics[width=\columnwidth,height=10cm]{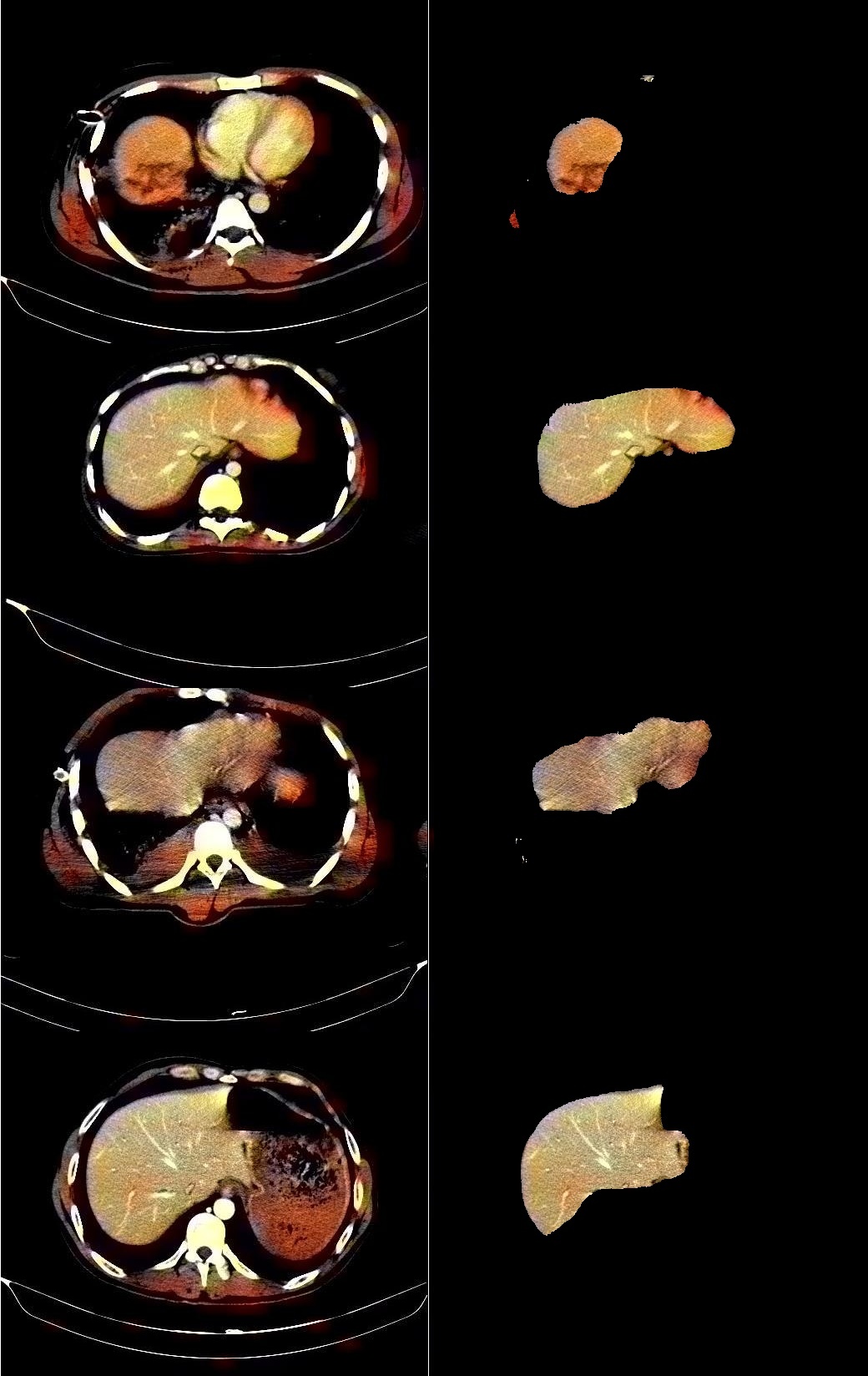}
   \caption{Applying model trained by 3DIRCAD for liver segmentation to Rajaee dataset.}
\end{figure}

\begin{figure}
\includegraphics[width=\columnwidth,height=10cm]{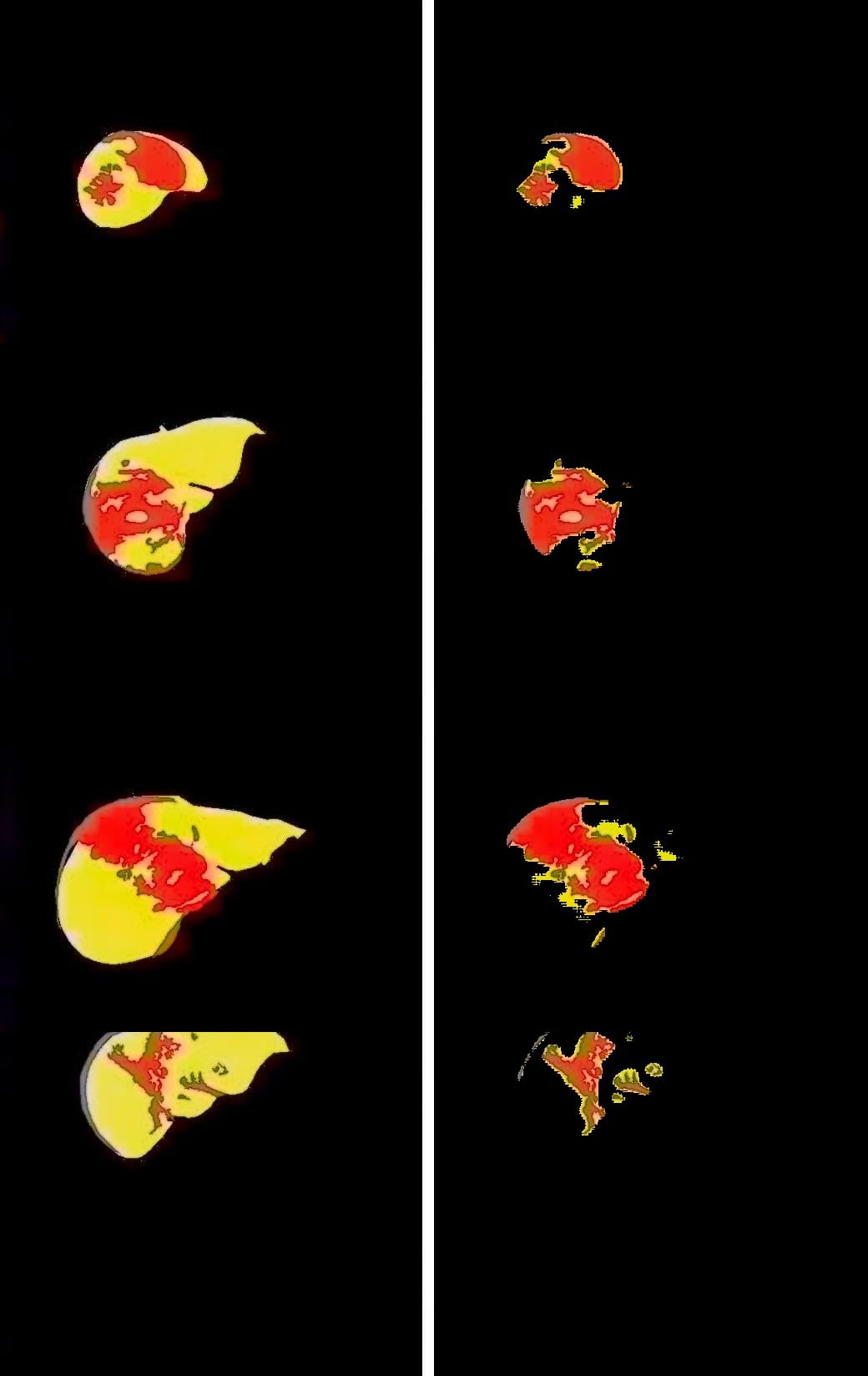}
   \caption{Applying model trained by 3DIRCAD for liver tumor or cyst to Rajaee dataset for laceration.}
\end{figure}

\begin{figure}
\includegraphics[width=\columnwidth,height=4cm]{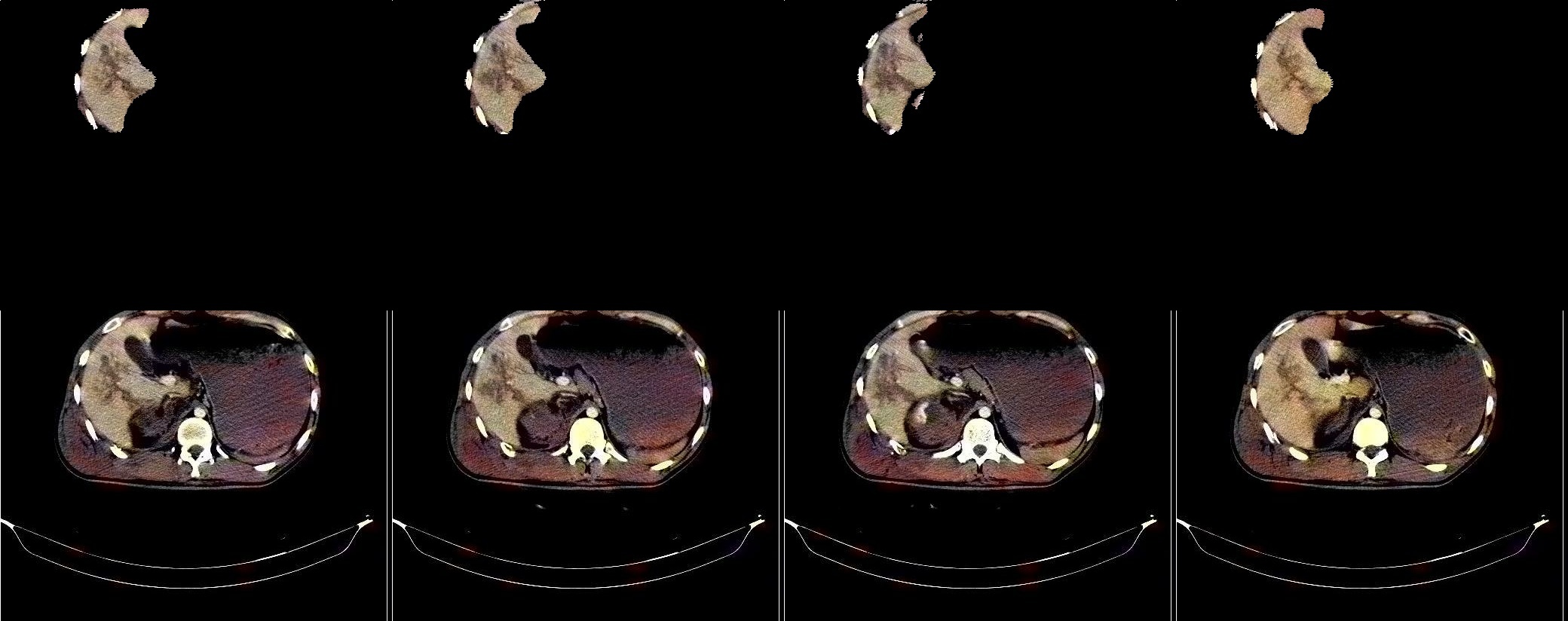}
   \caption{Applying model trained by Rajaee for liver segmentation to unseen data from Rajaee dataset.}
\end{figure}

\begin{figure}
\includegraphics[width=\columnwidth,height=4cm]{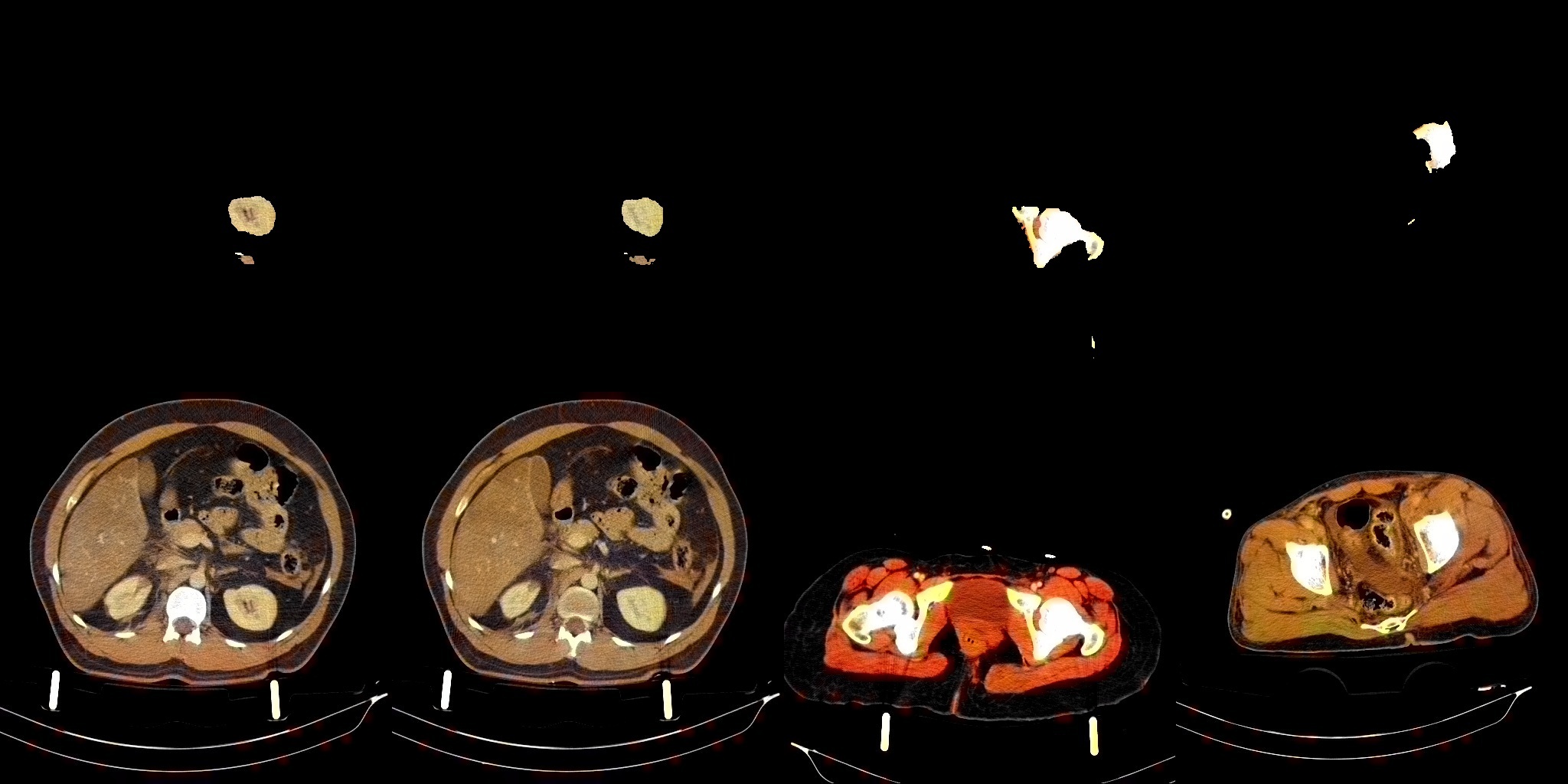}
   \caption{Applying model trained by 3DIRCAD for left kidney to unseen data from Rajaee dataset.}
\end{figure}

\begin{figure}
\includegraphics[width=\columnwidth,height=4cm]{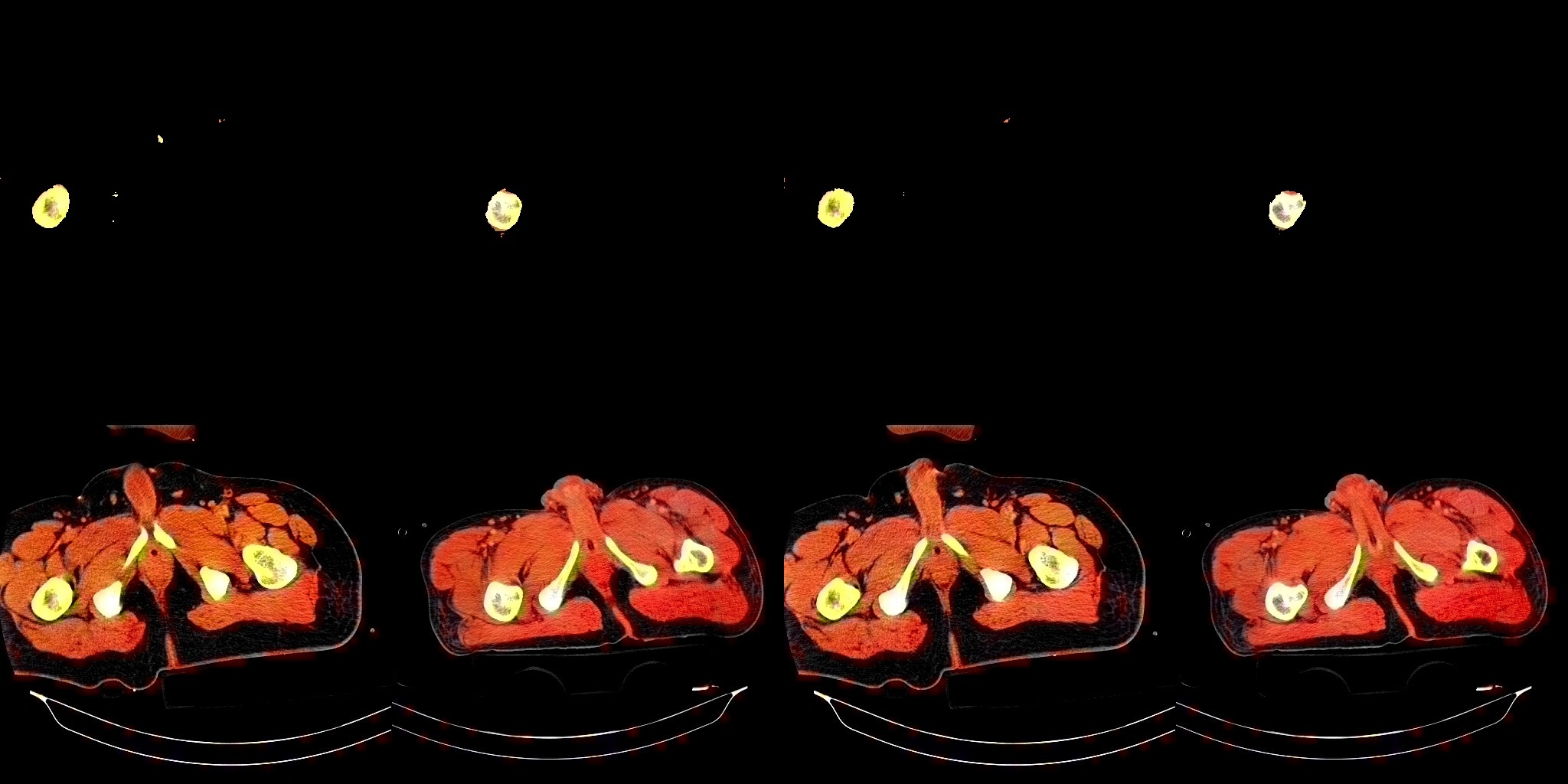}
   \caption{Applying model trained by 3DIRCAD for right kidney to unseen data from Rajaee dataset.}
\end{figure}

\begin{figure}
\includegraphics[width=\columnwidth,height=4cm]{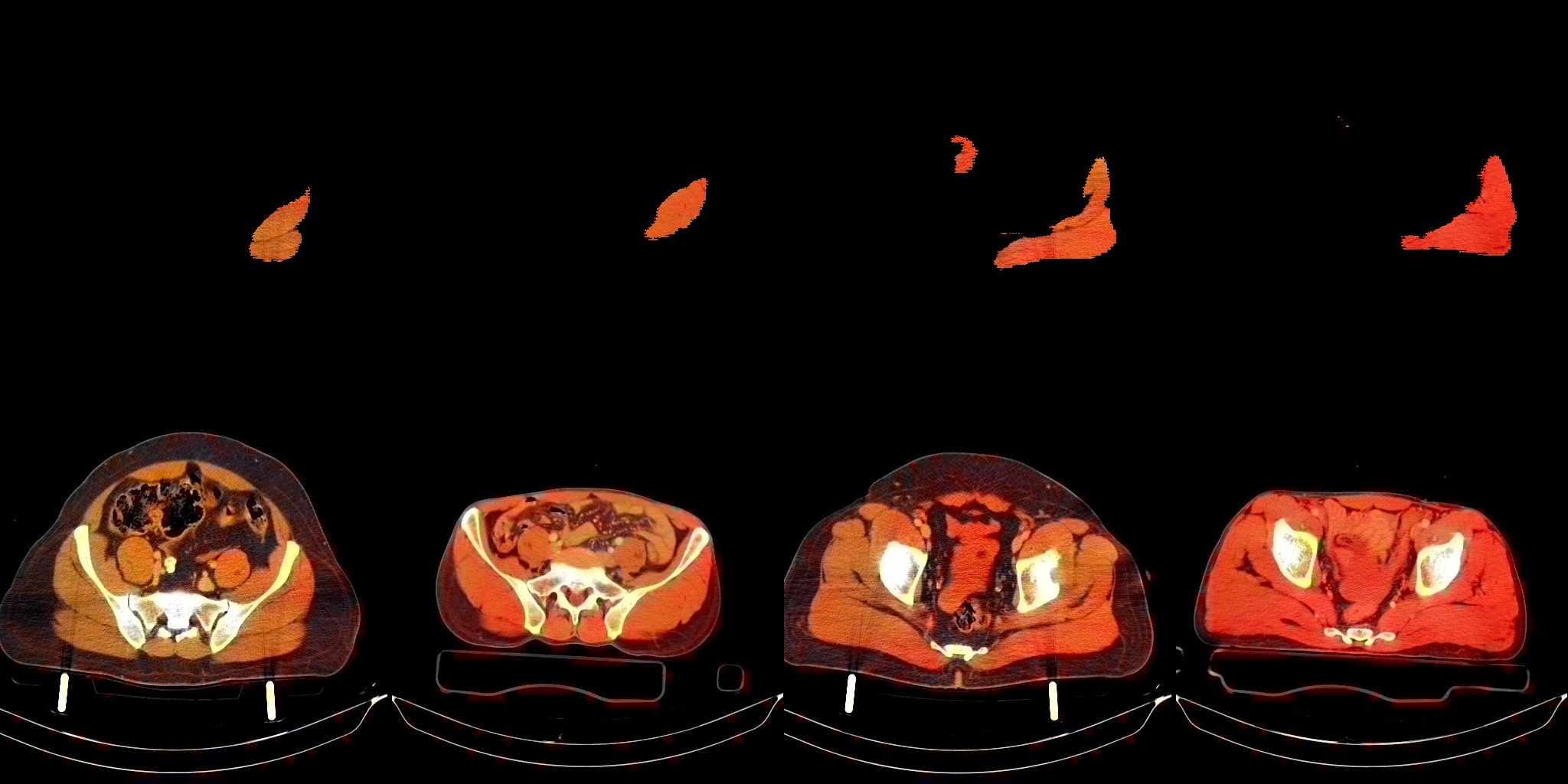}
   \caption{Applying model trained by 3DIRCAD for spleen to unseen data from Rajaee dataset.}
\end{figure}

\section{Conclusion and Further Developments}

In a hope to address automated detection of multiple-organ lacerations. 
 We presented the results of automated segmentation of lacerated organs and bleeding areas that could
help identify and differentiate injuries across multiple organs,
providing a comprehensive assessment essential for effective
treatment planning. This capability is particularly beneficial
in scenarios where radiologists may be overwhelmed or when
their expertise is not immediately available. 

The study introduced a highly efficient and accurate decision support system for diagnosing liver trauma using the Pix2Pix GAN model for CT scan image analysis. Achieving a Dice score of up to 97\% for liver bleeding and 93\% for liver laceration detection, this system significantly surpasses existing methods by delivering more precise and quicker segmentation. Integrating this system into clinical settings could enhance trauma management by providing rapid diagnostic assessments, improving treatment response times, and potentially reducing healthcare costs associated with extended diagnostic procedures.

In addition, we presented extensions to other vital organs involved in abdominal traumas, such as the kidneys and spleen. Expanding the model to recognise and accurately segment injuries to these organs could transform the system into a comprehensive abdominal trauma assessment tool.  This expansion could significantly improve the overall effectiveness of trauma care by enabling rapid, simultaneous assessments of multiple injuries, thereby facilitating more informed and timely surgical decisions. Additionally, by covering more organs, the system could be used in a wider range of emergency scenarios, increasing its utility and impact in clinical settings.

This research sets a foundation for significant advancements in medical imaging and trauma diagnosis, potentially transforming emergency medical services with AI-driven solutions.

\textbf{ }

\textbf{Ethical Approval and Consent to Participate:} This study has received ethical approval from the Shiraz University of Medical Sciences Ethics Committee. Approval number: IR.SUMS.REC.1402.101. Due to the retrospective nature of the study and anonymized data, written informed consent for participation was waived by the ethics committee.

\textbf{ }

\textbf{Consent for Publication:} Due to the retrospective design of the study and the anonymization of patient data, the ethics committee waived the need for written informed consent for publication.


\bibliographystyle{model1-num-names}

\bibliography{cas-refs}

\end{document}